# CONAN: A Python package for modeling lightcurve and radial velocity data of exoplanetary systems

**Babatunde Akinsanmi** [1,¶], **Monika Lendl** [1], **and Andreas Krenn** [2]

**1** Observatoire astronomique de l'Université de Genève, chemin Pegasi 51, 1290 Versoix, Switzerland **2** Space Research Institute, Austrian Academy of Sciences, Schmiedl-strasse 6, A-8042 Graz, Austria **¶** Corresponding author

## Summary

CONAN (COde for exoplaNet ANalysis) is an open-source Python package to perform comprehensive analyses of exoplanetary systems. It provides a unified Bayesian framework to simultaneously analyze diverse exoplanet datasets to derive global system parameters. CONAN allows to consistently model photometric transit light curves, occultations, phase curves, and radial velocity measurements, while detrending each dataset with any combination of parametric, sinusoidal, Gaussian Processes, and spline models.

## Statement of need

Detecting and characterizing exoplanets, planets orbiting stars other than our Sun, is a major focus area of astronomical research. This endeavor increasingly relies on heterogeneous datasets spanning multiple epochs, observing techniques, instruments, and wavelengths. As such, robustly estimating the physical and orbital properties of planets routinely requires the simultaneous modeling of the different signals, while dealing with the unique systematics of each instrument and the time-dependent impact of stellar variability.

CONAN is designed to address these needs with a number of key features:

- Multi-dataset analysis: Seamless analysis of combined lightcurve (LC) and radial velocity (RV) datasets from various instruments.
- Multiplanet support: Simultaneous fit to multiple planets in a single system.
- Comprehensive photometric modeling: Robust modeling of transits, occultations, and phase curves, including effects such as ellipsoidal variations and Doppler beaming (see Model definition).
- Modeling time- and wavelength-dependent signals: Analysis of light curves with transit timing variations (TTVs) and transit depth variations (transmission spectroscopy).
- Flexible baseline and noise modeling: Selection of one or combination of polynomial, sinusoidal, multi-dimensional Gaussian Processes (GP), and spline functions for data detrending.
- Extensible and customizable modeling: Users can easily incorporate new LC and RV models or modify default ones to suit specific needs, e.g., modeling the transit of non-spherical planets, Rossiter–McLaughlin signals, or even non-planetary signals.
- Robust Bayesian inference: Parameter estimation and/or model comparison via nested sampling with `dynesty` (Speagle, 2020) or Markov-Chain Monte Carlo sampling with emcee(D. Foreman-Mackey et al., 2013).

- Derivation of priors on limb darkening coefficients: Derive priors for the quadratic limb darkening coefficients (Kipping, 2013) from the stellar parameters using `ldtk` (Parviainen & Aigrain, 2015).

- Automated selection of parametric model parameters: Uses the Bayesian Information Criterion to suggest best combination of cotrending basis vectors for the analysis of time-series data.

- Science data download: Built-in support for downloading data from various instruments (including TESS, CHEOPS, Kepler, and K2) and also system parameters from NASA Exoplanet Archive.

- Quick result visualization and manipulation: Instant plot of the best-fit model and a result object that can be easily manipulated for customized analysis. A sample of an instant plot obtained from a `CONAN` fit is shown in Figure 1.

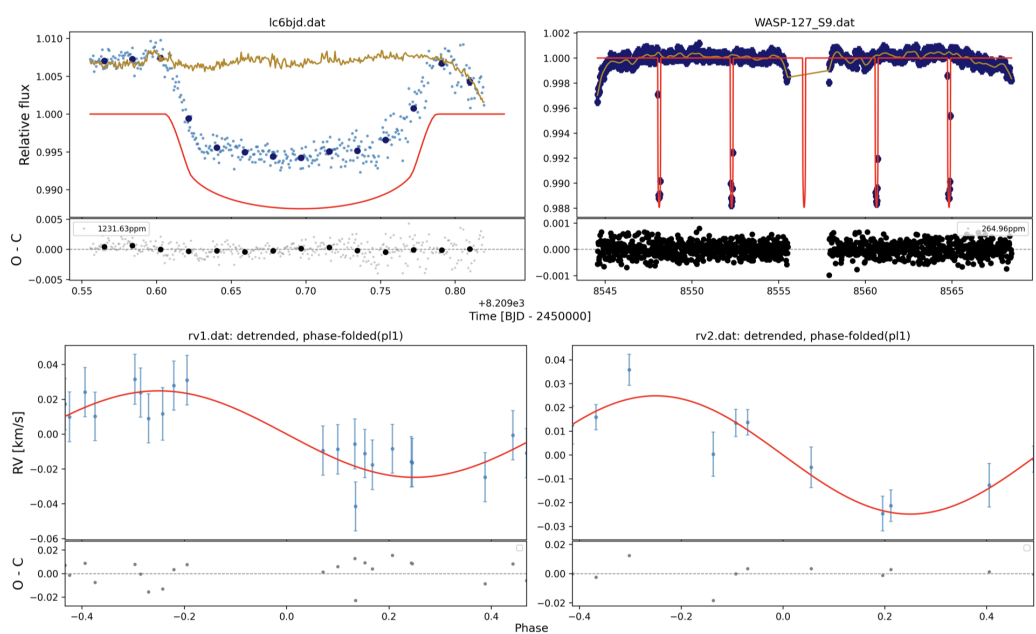

Figure 1: An example of joint fit to transit and RVs of WASP-127 b. The top panels show the best-fit models to the ground-based (left) and TESS (right) undetrended lightcurves. The transit model is shown in red, while the detrending baseline model is shown in gold (left: parametric model; right: GP). The bottom panels show the best-fit RV model overplotted on the detrended and phasefolded data. The details of the fit can be found in the online documentation.

`CONAN` was first introduced in (Lendl et al., 2017), and has been widely used in 15 peer-reviewed publications (Petit dit de la Roche et al., 2024; e.g., Psaridi et al., 2023; Seidel et al., 2025) with a total of 419 citations.

There are similar tools to `CONAN` for performing joint fit to exoplanet data, each with its own strengths and limitations. Some of these include `Juliet`(Espinoza et al., 2019), PyOrbit (Malavolta, 2016), exoplanet(Daniel Foreman-Mackey et al., 2021), Pyaneti(Barragán et al., 2019), ExoFAST(Eastman et al., 2013). One of the main strengths of `CONAN` compared to these tools is its capability to fit a wider variety of planetary signals. None of these publicly available tools can model full-orbit phasecurves of exoplanets using different phase functions. Notably, `CONAN` allows the user to define the custom model they would like to use in fitting the data, opening up practically unlimited use cases for `CONAN`. Additionally, `CONAN`'s capability to automatically select the best cotrending basis vectors for a dataset makes it especially well-suited to modeling extensive sets of ground-based observations.

The full documentation can be accessed at https://conan-exoplanet.readthedocs.io

# Acknowledgements

We would like to thank Angelica Psaridi, Hritam Chakraborty, Dominique Petit dit de la Roche, and Adrien Deline for their help in testing CONAN for several use cases. BA and ML acknowledge the support of the Swiss National Science Foundation under grant number PCEFP2_194576. CONAN makes use of several publicly available packages such as emcee (D. Foreman-Mackey et al., 2013), dynesty (Speagle, 2020), Astropy (Astropy Collaboration et al., 2022), celerite (D. Foreman-Mackey et al., 2017), spleaf (Delisle et al., 2020), lightkurve (Lightkurve Collaboration et al., 2018), numpy (Harris et al., 2020), ldtk (Parviainen & Aigrain, 2015). We thank the developers of these packages for their work.